\documentclass[epj,referee]{svjour}
\usepackage{graphics}
\begin{document}
\title{0-$\pi$ oscillations in nanostructured Nb/Fe/Nb Josephson junctions}
\author{Samanta Piano \inst{1,2} \and J. W. A. Robinson\inst{1} \and Gavin
Burnell\inst{3} \and Mark G. Blamire\inst{1}
}                     
%
%
\institute{
  \inst{1} Department of Materials Science, University of Cambridge - Pembroke Street, Cambridge CB2 3QZ, UK\\
  \inst{2} Physics Department, CNR-Supermat Laboratory, University of
  Salerno - Via S. Allende, 84081 Baronissi (SA), Italy\\
  \inst{3} School of Physics and Astronomy, E.C. Stoner Laboratory,
University of Leeds - Leeds, LS2 9JT, UK}
\date{Received: date / Revised version: date}
%
\abstract{ The physics of the $\pi$ phase shift in ferromagnetic
Josephson junctions may enable a range of applications for
spin-electronic devices and quantum computing. We investigate
transitions from ``0'' to ``$\pi$'' states in Nb/Fe/Nb Josephson
junctions by varying the Fe barrier thickness from $0.5$ nm to $5.5$
nm. From magnetic measurements we estimate for Fe a magnetic dead
layer of about $1.1$ nm. By fitting the characteristic voltage
oscillations with existing theoretical models we extrapolate an
exchange energy of $256$ meV, a Fermi velocity of $1.98 \times 10^5$
m/s and an electron mean free path of $6.2$ nm, in agreement with
other reported values. From the temperature dependence of the
$I_CR_N$ product we show that its decay rate exhibits a nonmonotonic
oscillatory behavior with the Fe barrier thickness.
} 
\maketitle
\section{Introduction}
\label{intro} In recent years a great interest has grown in the
implementation of quantum computing \cite{NielsenChuang}, where
information is stored in two-level quantum systems (qubits). Many
theoretical and experimental advances have been achieved towards the
physical realisation of qubits and, in this respect, solid-state
implementations appear as good candidates thanks to their potential
scalability which is partly enabled by the development of nano-scale
technology \cite{wallraffreview}. In particular, ferromagnetic
$\pi$-junctions have been proposed as coherent two-state quantum
systems \cite{YamaPRL}. In a Superconductor/Ferromagnet (S/F) system
the superconducting order parameter penetrates into the F layer. As
a consequence of exchange splitting of the spin-up and spin-down
electrons in the ferromagnetic sub-bands the superconducting order
parameter oscillates as a function of depth in the F-layer which
causes its sign to change periodically
\cite{Aarts,Kontos2001,BuzdinReview}. A positive order parameter
corresponds to a ``0''-state, while a negative order parameter to a
``$\pi$''-state.  To realise a practical quantum device it is
necessary to control these two states; this is easily achieved in
S/F/S systems incorporating weak \cite{Kontos2002} and strong
\cite{Blum,Bell,PRL} F barriers and hence such systems have been
identified as suitable quantum-electronic devices based on
$\pi$-shift technology \cite{YamashitaPRB}. These systems exhibit
oscillations of the critical current with F layer thickness
equivalent to a change of phase from $0$ to $\pi$-states. The
principle goal in this research field is to understand the physics
of superconductor based quantum technology and to evaluate potential
material systems, while aiming at reducing the size of the employed
heterostructures towards the nano-scale.

In this paper, we present the first study of the thickness
dependence of the critical current oscillations in
niobium/iron/niobium (Nb/Fe/Nb) Josephson junctions. Magnetic
measurements as a function of the Fe layer thickness ($t_{Fe}$) have
been made so that an estimate of the magnetic dead layer of our Fe
barrier layer can be obtained. The current \emph{vs} voltage
characteristics have been measured for different thicknesses of the
Fe and show periodic transitions from $0$-states to $\pi$-states
from nodes in the $I_c R_N (t_{Fe})$ relation, where $I_c$ is the
critical current of the device and $R_N$ is the normal-state
resistance. From a comparison between experimental data and two
theoretical models we have estimated the exchange energy and the
Fermi velocity of the Fe barrier, which are found to be in good
agreement with the expected values from literature. With a simple
model we estimate the mean free path of Fe verifying that the clean
limit condition is fulfilled. From the study of the temperature
dependence of the $I_CR_N$ product we have shown that its decay rate
exhibits a nonmonotonic oscillatory behavior with the Fe barrier
thickness. Our analysis reveals that Fe, for its small magnetic dead
layer and high exchange energy, is well-suited for the
implementation into nano-scale superconductor based quantum
electronics.

\begin{figure}\centering
\resizebox{0.9\columnwidth}{!}{%
  \includegraphics{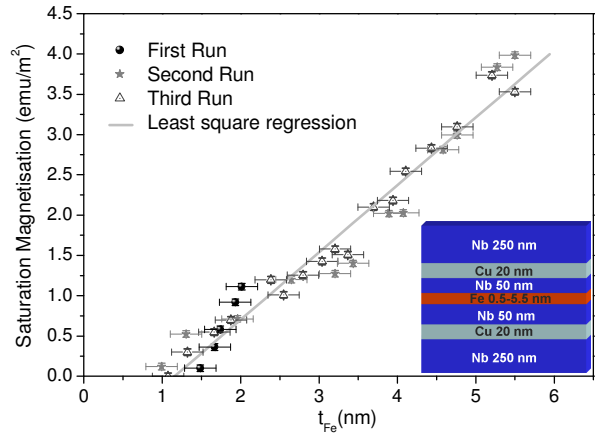}
}
\caption{(Colour online) Saturation magnetisation \emph{vs.} Fe
barrier thickness for three deposition runs. The measurements were
taken at room temperature. From the linear fit the magnetic dead
layer is extrapolated. Inset: illustration of our multilayers.}
\label{deadlayer}       
\end{figure}

\section{Magnetic properties of Nb/Fe/Nb}
Our heterostructures consist of Nb (250 nm) / Fe (0.5 nm - 5.5 nm) /
Nb (250 nm). To assist subsequent processing in a focused ion beam
(FIB) microscope, 20 nm of Cu was deposited inside the outer Nb
electrodes, but 50 nm away from the Fe barrier. We remark that the
20 nm of Cu is a thickness smaller than its coherence length, so it
is completely proximitised into the Nb, and does not affect the
transport properties of the Josephson junction. The heterostructures
are deposited by d.c. magnetron sputtering in an Ar plasma at 1.5 Pa
on $10\times 5$ mm silicon (100) substrates coated with a 250 nm
thick oxide layer on the surface. Our deposition system is equipped
with a rotating holder which can move under three magnetrons and
allows the loading of more than twenty substrates in one run. In
this way our system permits, knowing the relation between the
deposition rate and the speed of the rotating holder, to deposit
{\it in-situ} and in a single run, different samples with various Fe
barrier thicknesses. From x-ray reflectivity measurements, we have
verified the control of Fe barrier thickness variation $t_{Fe}$ to
be within an accuracy of $\pm 0.2$ nm \cite{NewPRB}.

To investigate the magnetic properties of our devices we have
measured, using a vibrating sample magnetometer at room temperature,
the magnetic moment per unit surface area of the films as a function
of Fe barrier thickness. The saturation magnetisation was measured
for three different deposition runs (see Fig. \ref{deadlayer}),
where for each deposition we have obtained similar saturation
magnetisations which also confirms our control of the Fe barrier
thickness. By extrapolating the least-squares fit of this data we
have estimated that the Fe has a magnetic dead layer $t_D \simeq
1.1$ nm. The thickness of the magnetic dead layer is one of the
central points into the realisation of quantum devices because it
marks the lower limit of the ferromagnetic barrier thickness for
which it exhibits magnetic moment. We remark that the scale of the
estimated Fe barrier magnetic dead layer is low; however, there is a
need for deposition optimisation to minimise the scale of the
magnetic dead layer so that better control and tunability of the Nb
phase-state is possible. The issues associated with a magnetic dead
layer are a particular problem with the strong ferromagnetic
barriers because of the very thin layers that have to be deposited
in order to exploit or even see $\pi$-shift physics and hence,
improved accessibility and knowledge of this dead layer is important
if strong ferromagnetic materials are to be incorporated into
$\pi$-shift based technologies.

\begin{figure}\centering
\resizebox{0.9\columnwidth}{!}{%
  \includegraphics{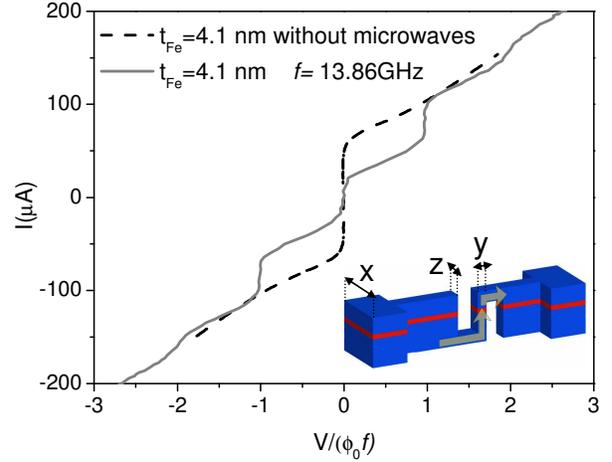}
}
\caption{(Colour online) $I$ \emph{vs.} $V/(\phi_0 f)$, where
$\phi_0=h/2e$, for a Josephson junction with an Fe barrier thickness
of 4.1 nm (black dash line) and the same device with 13.86 GHz
excitation of microwaves (red line) at T$=4.2$ K. Inset: an
illustration of a focused ion beam processed device where $x\approx
4\mu$m, $100 \leq y \leq 500$ nm and $100 \leq z \leq 500$ nm. The
current path is also shown.}\label{shapiro}  
\end{figure}

\section{Realization of nanostructured Josephson junctions and electrical characterization}
The Josephson junctions were fabricated in three steps: (i)
patterning of films using optical lithography to define micron-scale
tracks and contact pads. Our mask permits 14 devices per sample
which allows us to measure numerous devices and to derive good
estimates of important parameters, like $I_c R_N$; (ii) broad beam
Ar ion milling ($3$ mAcm$^{-2}$, $500$ V beam) to remove unwanted
material from around the mask pattern which leaves $4 \mu$m tracks
for subsequent FIB work; (iii) FIB etching of micron-scale
tracts\cite{Chrisnano} to achieve vertical transport with a device
area (yz) in the range $0.2$ $\mu$m$^2$ to $1.1$ $\mu$m$^2$. We show
a labeled illustration of a device in the inset of Fig.
\ref{shapiro}.

Transport measurements were made in a custom made liquid He dip
probe with a heating stage and microwave antenna fitted.
Measurements were made with a standard lock-in technique. The
critical current $I_c$ was found from the differential resistance as
the point where the differential resistance increases above the
value for zero bias current. $R_N$ was measured using a quasi-dc
bias current of 3-5 mA, this enabled the nonlinear portion of the
I-V curves to be neglected, but was not large enough to drive the Nb
electrodes into a normal state.

In Fig. \ref{shapiro} we show $I_c R_N$ \emph{vs.} $V/\phi_0 f$ for
a device with an Fe barrier thickness of $\simeq0.8$ nm with and
without microwaves applied, where $\phi_0$ is the flux quantum given
by $h/2e$ and $f$ is the applied microwave frequency ($f=14.01$
GHz). Constant voltage Shapiro steps appear due to the
synchronisation of the Josephson oscillations on the applied
excitation \cite{Barone} at voltages equal to integer multiples of
$hf/2e=0.029$ meV.

\section{Critical current oscillations}
\begin{figure}\centering
\resizebox{0.9\columnwidth}{!}{%
  \includegraphics{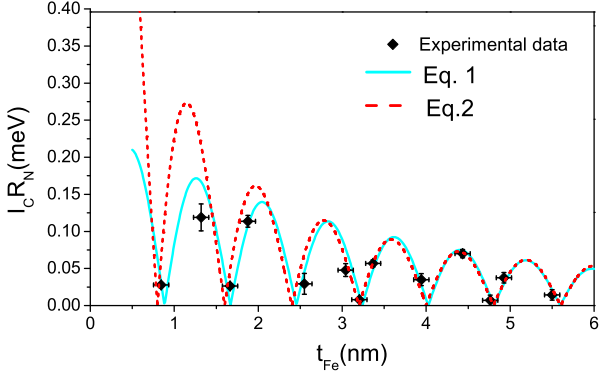}}
\caption{(Colour online) $I_cR_N$ oscillations as a function of the
thickness of the iron barrier
  (dot) with the theoretical fit (line).}\label{oscillations}
\end{figure}

For each sample we have measured, from the $I-V$ characteristics, at
least three junctions from which the critical current ($I_c$) and
the normal resistance ($R_N$) of a device is extracted. The $I_c
R_N$ product is then plotted as a function of Fe barrier thickness
and reveals $I_c R_N(t_{Fe})$ decaying with multiple oscillations in
the clean limit up to $5$ nm (see Fig.\ref{oscillations}).

The experimental data have been modeled with the following
formula\cite{eq1}:
\begin{equation}
I_c R_N = I_c R_N (t_0)\left\vert
\frac{\sin\frac{t_{Fe}-t_1}{\xi_2}}{\sin
\frac{t_1-t_0}{\xi_2}}\right \vert
\exp\left(\frac{t_0-t_{Fe}}{\xi_1}\right), \label{general}
\end{equation}
where $t_{1}$ is the Fe thickness corresponding to the  first
minimum in $I_c R_N$, $I_c R_N (t_0)$ is the first experimental
value of $I_c R_N(t_{Fe})$, and $\xi_1$ and $\xi_2$ are the two
fitting parameters. We notice that the experimental data are in good
agreement with this theoretical model, and from the fit we obtain
$\xi_1=3.8$ nm and $\xi_2=0.25$ nm and a period of the oscillations
is $t=1.6$ nm. Eq. (\ref{general}) is a general formula that applies
to the clean and dirty limits; however, in the clean limit $\xi_2=
v_F \hbar/2 E_{ex}$. By assuming the Fermi velocity of Fe is
$v_F=1.98 \times 10^5$ m/s, as reported in
literature\cite{literature}, we calculate the exchange energy of the
Iron: $E_{ex}= \hbar v_F/2t\approx 256 {\ \rm meV}$.

To confirm that our oscillations are in the clean limit (meaning
that the considered Fe thickness is always smaller than the Fe mean
free path), we have modeled our data with a second simplified
formula which holds only in this limit \cite{eq2}:
\begin{equation}
I_c R_N \propto  \frac{\mid \sin(2 E_{ex} t_F /\hbar v_f) \mid}{2
E_{ex} t_F /\hbar v_f}, \label{cleanlimit}
\end{equation}
where, in this case, $E_{ex}$ and  $v_F$ are the two fitting
parameters. From the theoretical fit we extrapolate $E_{ex}=256$ meV
and $v_F=1.98 \times 10^{5}$ m/s.

This model is fitted using exactly the same values for the
parameters as those obtained from the previous model given by Eq.
\ref{general}. Both models are thus consistent with one another and
show an excellent fit to our data, which exhibit multiple
oscillations of $I_c R_N$ in a small range of Fe barrier thickness.
In Fig. \ref{oscillations} we show the experimental data with the
two theoretical fits. The magnetic dead layer has not been
subtracted, as for T$=4$K it is expected to be significantly lower
than $1.1$ nm. This is an important feature towards the realisation
of integrated nanostructured quantum electronic devices based on
$\pi$-phase shift in Nb/Fe/Nb Josephson junctions.

With a simplified model that is obtained solving the linearised
Eilenberger equations \cite{Gusakova} we can estimate the mean free
path of Fe. The general formula is:

\begin{equation}
\ \tanh \frac{L}{\xi_{eff}}=\frac{\xi_{eff}^{-1}}{\xi_0^{-1} +
L^{-1} + i \xi_H^{-1}}, \label{Born}
\end{equation}
where $\xi_{eff}$ is the effective decay length given by
$\xi_{eff}^{-1} = \xi_1^{-1} + i \xi_2^{-1}$, $\xi_o$ is the
Ginzburg-Landau coherence length and $\xi_H$ is a complex coherence
length. In the clean limit $1+L \xi_0^{-1}>>($max$\{ \ln (1 + L
\xi_0^{-1}), \ln(L \xi_H^{-1})\}$)/2. The solution of this equation
gives:  $\xi^{-1}_1 = \xi_0^{-1} + L^{-1}$, $\xi_0=(v_F \hbar)/(2
\pi T_c k_B)$ and $\xi_2 = \xi_H$.

\begin{figure}\centering
\resizebox{0.9\columnwidth}{!}{%
  \includegraphics{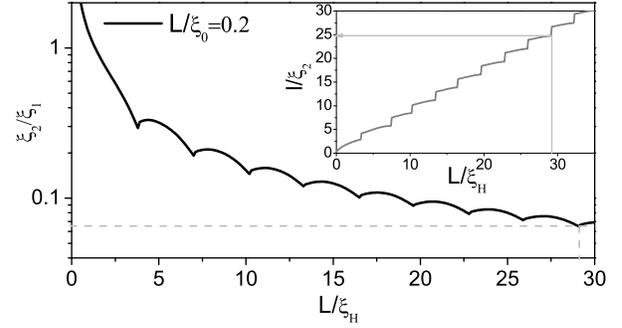}}
\caption{Dependence of $\xi_2/\xi_1$ with the inverse magnetic
length; inset: $L/\xi_H$ \emph{vs} $L/\xi_H$ to estimate the mean
free path, $L$.}\label{Febornfit}
\end{figure}

In our case the ratio $\xi_2/\xi_1 \simeq 0.06$, and so assuming
$L/\xi_0$=$0.2$ we can extrapolate from the graphical solution a
value of $L/\xi_H \simeq 29$ (Fig.\ref{Febornfit}). Considering the
curve $L/\xi_2$ \emph{vs} $L/\xi_H$ (inset in Fig.\ref{Febornfit})
we obtain a value of $L/\xi_2$ to be $\simeq25$ and hence, we
estimate the mean free path of Fe to be $\simeq6.2$ nm. With this
analysis we remark that the condition that $I_c R_N(d_{Fe})$ is
within the clean limit only is fulfilled.

\begin{figure}\centering
\resizebox{0.9\columnwidth}{!}{%
  \includegraphics{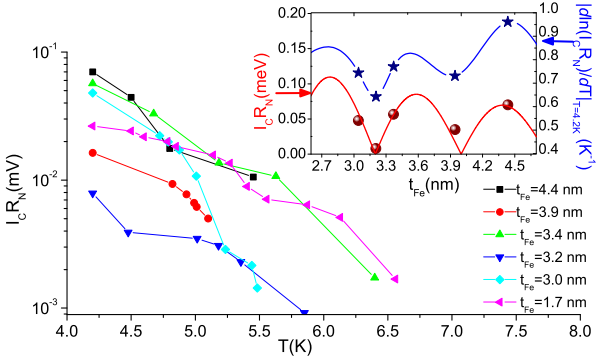}}
\caption{(Colour online) Temperature dependence of the $I_CR_N$ for
  different thicknesses of the Fe layer (logarithmic scale). Inset: oscillations of the decay
  rate $\alpha_T(T_0=4.2{\rm K})$ of $I_CR_N$, Eq.~(\ref{alpha}), with the temperature [stars; the blue line is a guide to the eye],
  as compared with the oscillations of the $I_CR_N$ product itself [spheres;
  the red line is the fit given by Eq.~(\ref{general})] in a window of Fe thicknesses.}\label{temp}
\end{figure}

We have followed the temperature dependence of the $I_CR_N$ product
for different thicknesses of the Fe barrier, as shown in
Fig.\ref{temp}. We note that, for each Fe barrier thickness, the
Josephson junction resistance remains approximately constant. On the
other hand, the critical current, and hence the $I_CR_N$ product,
quickly decreases with increasing the temperature. It is interesting
to notice how the rate at which the critical current drops to zero
exhibits a nonmonotonic dependence on the Fe layer thickness. We can
define the relative decay rate $\alpha_T$ of the $I_CR_N$ product
with the temperature as in Ref. \cite{Blum}, namely
\begin{eqnarray}\label{alpha}
\alpha_T(T_0)&=&\left|\frac{{\rm d} \ln[I_CR_N(T)]}{{\rm
d}T}\right|_{T=T_0} \\ &=& \left|\frac{{\rm d} I_CR_N(T)}{{\rm
d}T}\right|_{T=T_0} \frac{1}{I_CR_N(T_0)}\,, \nonumber
\end{eqnarray}
 calculated at $T_0=4.2$ K. In the inset of  Fig.\ref{temp} we plot $\alpha_T$ as a
function of $t_{Fe}$ for five consecutive thicknesses; we observe
oscillations of $\alpha_T$ in phase with the oscillations of the
$I_CR_N$, in other words the smaller $I_CR_N$ decays to zero with
increasing $T$ more slowly than the greater $I_CR_N$.

\section{Conclusions}

In summary, we have fabricated Nb/Fe/Nb nano-scale Josephson
junctions and investigated the effect of varying the Fe barrier
thickness on $I_c R_N$. From magnetic measurements the Fe barriers
were found to have a magnetic dead layer of $\simeq 1.1$ nm. From
$I-V$ curves we have measured the Josephson critical current ($I_c$)
and the normal-state resistance $R_N$ so that oscillations in $I_c
R_N$ as a function of the Fe barrier thickness could be studied. In
agreement with two theoretical models we have fitted our data and
estimated the exchange energy of Fe to be $256$ meV and the Fermi
velocity of Fe to be $1.98 \times 10^5$ m/s. Furthermore, we have
applied a simple theoretical model that involves solving the
linearised Eilenberger equations so that an estimate of the Fe
electron mean free path could be made ($6.2$ nm). A value of $6.2$
nm confirms that $I_c R_N(t_{Fe})$ oscillations are fully in the
clean limit giving rise to a maximised $I_c R_N$ product in the
$\pi$-state. For different Fe barrier thicknesses we have shown that
the $I_CR_N$ product decreases with increasing temperature, and in
particular the decay rate presents the same oscillatory behavior as
the critical current, this original feature seems common to strong
ferromagnetic $\pi$ junctions \cite{NewPRB}, although no explanation
for this can yet be given. With this work we have shown that it is
possible to fabricate nanostructured Nb/Fe/Nb $\pi$-junctions with a
small magnetic dead layer  of $1.1$ nm and with a high level of
control over the Fe barrier thickness variation. The estimated
exchange energy of Fe is close to bulk Fe implying that the Fe is
clean and S/F roughness is minimal. Some interfacial diffusion of Fe
into Nb could account for the slight suppression of $E_{ex}(Fe)$ and
the magnitude of the magnetic dead layer. We conclude that Fe
barrier S/F/S Josephson junctions are viable structures in the
development of superconductor-based quantum electronic devices. The
electrical and magnetic properties of Fe are well understood and are
routinely used in the magnetics industry, therefore Nb/Fe/Nb
multilayers can readily be used in controllable two-level quantum
information systems.

%

%
%

\end{document}